\documentclass[twocolumn,final,showpacs,superscriptaddress,aps]{revtex4}

\usepackage{graphics}
\usepackage{graphicx}

\begin{document}
\title{Micron-sized atom traps made from magneto-optical thin films}

\author{S. Eriksson}
\affiliation{Blackett Laboratory, Imperial College, London SW7 2BW, 
United Kingdom}
\author{F. Ramirez-Martinez}
\affiliation{Blackett Laboratory, Imperial College, London SW7 2BW, 
United Kingdom}
\author{E.A. Curtis}
\affiliation{Blackett Laboratory, Imperial College, London SW7 2BW, 
United Kingdom}
\author{B.E. Sauer}
\affiliation{Blackett Laboratory, Imperial College, London SW7 2BW, 
United Kingdom}
\author{P.W. Nutter}
\affiliation{Department of Computer Science, University of Manchester, 
Manchester M13 9PL, 
United Kingdom}
\author{E.W. Hill}
\affiliation{Department of Computer Science, University of Manchester, 
Manchester M13 9PL, 
United Kingdom}
\author{E.A. Hinds}
\affiliation{Blackett Laboratory, Imperial College, London SW7 2BW, 
United Kingdom}

\begin{abstract}
We have produced magnetic patterns suitable for trapping and
manipulating neutral atoms on a $1\,\mu$m length scale.  The
required patterns are made in Co/Pt thin films on a silicon
substrate, using the heat from a focussed laser beam to induce
controlled domain reversal. In this way we draw lines and
``paint'' shaped areas of reversed magnetization with sub-micron
resolution. These structures produce magnetic microtraps above the
surface that are suitable for holding rubidium atoms with trap
frequencies as high as $\sim\,1$ MHz.
\end{abstract}

\maketitle

\section{Introduction}
\label{intro} The study of cold atom clouds in microscopic
magnetic traps and waveguides has recently become a vigorous field
of research\,\cite{OurReview,FolmanReview}. In particular, the
ability to confine and manipulate atoms above microstructured
surfaces - known as atom chips - holds great promise for
integrated atom optics and the realisation of new quantum devices.
Examples include miniature atom
interferometers\,\cite{Hinds2001,Haensel2001} and quantum
information processors\,\cite{Calarco2000}.

Most atom chip experiments to date have produced the trapping
fields by passing current through integrated wires. This approach
has two undesirable features. First, magnetic field noise is
generated by thermal fluctuation of the charges in the
wires\,\cite{Henkel1999,Rekdal2004}. This can change the internal
state of the atoms through magnetic dipole spin-flip
transitions\,\cite{Jones2003,Harber2003}, causing decoherence and
trap loss. Second, the current does not flow as intended within
the wires, but wanders from side to side\,\cite{Jones2004},
resulting in traps of uneven depth. An atom cloud cooled in such a
trap breaks into fragments when the available energy becomes
comparable with the amplitude of the noise in the trap depth
\cite{Fortagh2002,Leanhardt2002,Jones2004}. These imperfections
are detrimental for many practical applications.

As an alternative, our group has been working on microscopic
patterns of permanent magnetisation\,\cite{OurReview}, which offer
a way around the problems discussed above by avoiding currents and
by keeping the metal layers thin\,\cite{Rekdal2004,Jones2003}.
Previous experiments used
videotape\,\cite{Rosenbusch2000,Jones2003}, on which patterns with
spatial periods as small as $10\,\mu$m are conveniently written.
Future experiments will aim for controlled tunnelling of atoms
between one trap and another as a powerful way to prepare quantum
states of the atomic motion\,\cite{Bloch2002}. This will require
magnetic structures of order $1\,\mu$m in size, prompting us to
investigate Co/Pt thin films. Small structures also make traps so
strong that atom clouds can be compressed down to one dimension
for new experiments in the physics of quantum gases
\,\cite{Stoof2002,Olshanii2001}.

In this paper, we describe the fabrication of atomic microtraps
based on magneto-optically (MO) patterned Co/Pt thin films. When
combined with a suitable uniform external field, the magnetic
field above the patterned surface of the film creates local minima
where alkali atoms prepared in a weak-field-seeking state can be
confined\,\cite{OurReview}. The first attempt to do this was by
the group of P. Hannaford using TbFeCo films, but with these it
did not seem possible to achieve the desired control over the
domain boundaries\,\cite{Lau1999}. Here we show by contrast that
Co/Pt films are suitable for writing well-defined structures on
the required $1\,\mu$m size scale. The strong magnetisation of the
cobalt in conjunction with the small scale of the structures
produces large magnetic field gradients of order $10^{4}\,$T/m.
The magnetic field noise will be low because the metal film is
thin. This will allow trapped rubidium atoms to be held at micron
distances from the surface where the strong field gradient gives
trapping frequencies in excess of $1\,$MHz, corresponding to
exceedingly tight atom confinement.

The magnetisation of the film is normal to the surface, providing
the freedom to create arbitrary 2D patterns in the plane of the
film. This is a key difference between in-plane and perpendicular
magnetic media. The writing technique used in this work is based
on a standard method used in data storage applications and has a
similar aim, i.e. to produce small scale patterns. However, high
density data storage aims to achieve small isolated spots on the
film, whereas our aim is to produce larger, uniformly magnetised
regions with high constrast and small feature sizes along the
boundary. MO techniques have not previously been used for this
purpose. Here we present the writing of two patterns that are
fundamental building blocks for atom trapping and manipulation.
The writing process is fully reversible: any pattern can be erased
and re-written, and in principle this could even be done in the
presence of trapped atoms. This offers an advantage in comparison
with lithographically prepared atom chips, which cannot be
reconfigured.

In the next section we describe the preparation of Co/Pt
multilayer thin films and we summarise their magnetic properties.
Section\,\ref{sec:writing} describes the experimental apparatus
and the procedure for writing. Section\,\ref{sec:analysis} is
devoted to analysis of the patterns that were written and a
discussion of their application to atom chips. We conclude with a
summary of our results.

\section{The C\lowercase{o}/P\lowercase{t} multilayer thin films}
\label{sec:multilayers}

Magneto-optical films have been much studied for their application
to data-storage\,\cite{Mansuripur}. Recently, Co/Pt films such as
we are using here have attracted interest because they give a
strong optical Kerr rotation to blue light, which can have a small
spot size.  High resolution makes Co/Pt attractive for writing
microscopic atom traps and guides as well as for data storage.
Other relevant properties are a strong saturation magnetisation
and a large perpendicular anisotropy, leading to a large
coercivity and a very square hysteresis loop\,\cite{LinJMMM1991}.
The perpendicular anisotropy depends strongly on the
crystallographic orientation of the layers: a thick Pt base layer
increases the coercivity dramatically by establishing good
$\{111\}$ texture on which to deposit the multilayer. The
squareness of the hysteresis loop is optimised by having control
over the layer thicknesses: 0.2-0.4\,nm for the Co, corresponding
to 1-2 monolayers, and $\sim 1\,$\,nm for the Pt
layers\,\cite{Zeper1989and1991}.

\begin{figure}
\includegraphics*[width=0.95\columnwidth]{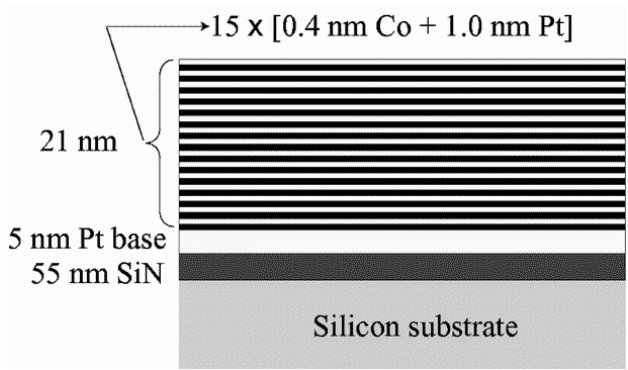}
 \caption{Layer structure used for the Co/Pt thin films (not to
scale).} \label{fig:film}
\end{figure}

The specific layer structure that we use is shown in
Fig.\,\ref{fig:film}. First, the silicon substrate has a
$55\,$nm-thick silicon nitride layer, which seems necessary to
achieve MO recording. In our first attempts on silicon we did not
have this insulating layer and we found that the 180\,mW of laser
power available for writing was not enough to make any observable
change in the film, either physical or magnetic. This problem is
rectified by the silicon nitride layer, which lowers the laser
power required for writing, presumably because it increases the
thermal resistance between the magnetic layers that must be heated
and the silicon wafer that acts as a heat sink.

The Pt and Co layers are laid down by vacuum deposition using an
electron beam source to evaporate the metals. The base pressure
before evaporation is $2\times 10^{-7}$ Torr, and the average
pressure during deposition rises to $3-4 \times 10^{-7}$ Torr. We
heat the substrate to a temperature of 200\,C in order to
establish adequate $\{111\}$ texture with a platinum base layer
that is only 5\,nm thick\,\cite{LinJMMM1991}. A thicker layer
gives better texture, but inhibits the MO writing by allowing the
heat to flow too quickly away from the laser spot. The
magneto-optical film itself consists of 15 bi-layers of
alternating Co (0.4\,nm) and Pt (1.0\,nm). The evaporation rates
are kept low at 10\,pm/s to ensure a good control of the layer
thicknesses.

\begin{figure}
\resizebox{0.45\textwidth}{!}{
  \includegraphics{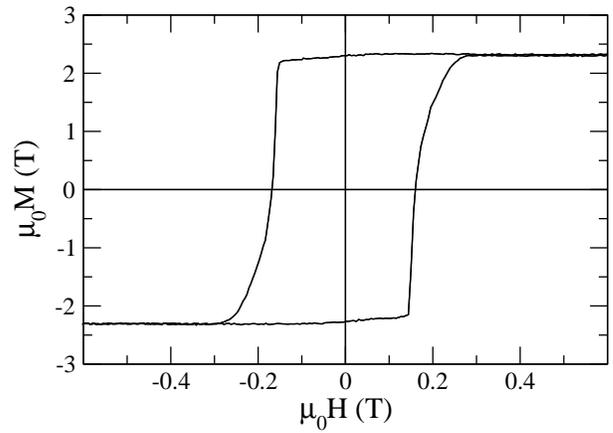}
} \caption{Alternating gradient force magnetometer measurement of
the hysteresis curve. Here the magnetic moment of the film is
expressed as an effective magnetisation within the 6\,nm thickness
of the Co layers.} \label{fig:hystloop}
\end{figure}

The magnetic properties were determined using an alternating
gradient force magnetometer, which yielded the hysteresis curve
shown in Fig.\,\ref{fig:hystloop}. The measured coercivity is
given by $\mu _0 H_c = 0.16$\,T. The measured magnetic moment can
be converted to an effective average magnetisation within the net
$6\,$nm thickness of the cobalt. Expressed in this way, the
magnetisation at saturation is given by $\mu_0 M_s = 2.3\,$T, some
30\% higher than that of bulk cobalt. This enhancement can be
attributed to additional magnetisation coming from polarized Pt
atoms near the Co layers\,\cite{LinJMMM1991}.

\section{Writing on the films}
\label{sec:writing}

The writing procedure depends upon the reduction of the coercivity
with increasing temperature\,\cite{writing}. The film is initially
driven into saturation in one direction normal to the surface by a
strong external magnetic field. This field is then reduced below
the room temperature coercivity and reversed. By illuminating the
sample locally with a focused laser beam, very small areas of the
film can be selectively heated and with an appropriate movement of
either the film or the laser beam, patterns of reversed
magnetization are created.

\subsection{Apparatus}

\begin{figure}
\includegraphics*[width=0.95\columnwidth]{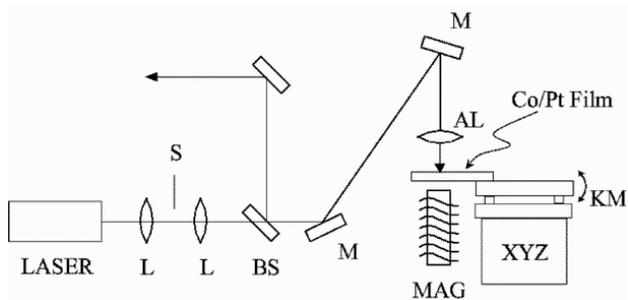}
\caption{Experimental setup for thermomagnetic writing.
Abbreviations: (L) lenses, (S) programmable shutter, (BS)
beam-splitter, (M) mirrors, (AL) aspheric lens, (MAG)
electromagnet, (KM) kinematic mount, and (XYZ) translation
stages.} \label{fig:writing}
\end{figure}

After using a permanent magnet to saturate it, the film is placed
between the pole faces of an electromagnet as shown schematically
in Fig.\,\ref{fig:writing}. This provides a reverse field of order
100\,mT for the purpose of writing patterns on the film. The laser
used for writing is a frequency doubled Nd:YVO laser operating at
532\,nm on the TM$_{00}$ transverse mode. The first lens L focuses
the light to a waist, where a mechanical shutter S is used to
switch the beam on and off. A second identical lens recollimates
it. Two mirrors M steer the beam to an aspheric lens AL with 0.55
numerical aperture and 4.4\,mm focal length that focuses the light
to a (calculated) waist of $0.6\,\mu$m, i.e. this is the radius at
which the intensity drops to $1/e^{2}$ of its peak value. The
Co/Pt film sits horizontally in a tray attached to a three point
kinematic mount KM, which is connected in turn to an $xyz$
translation stage.

Approximately $54\,\%$ of the light is reflected, allowing us to
position the film in the focal plane by translating it along the
vertical $z$ direction until the retro-reflected beam is
collimated. A beam-splitter BS picks off the back reflection for
this purpose. Since the Rayleigh length of the focussed light is
only $2\,\mu$m, this requires delicate adjustment, for which we
use a differential micrometer.

In order to write magnetic patterns, the film is moved
horizontally by a translation stage under computer control.
Initially, the plane of this $xy$ translation does not coincide
exactly with the  surface of the film, but through iterative
adjustments of the kinematic mount, we make the two planes
coincide so that the spot is always focussed on the surface. The
next two subsections describe how we use this setup to write
patterns suitable for atom trapping.

\subsection{Writing an array of lines}

\begin{figure} \resizebox{0.45\textwidth}{!}{
  \includegraphics{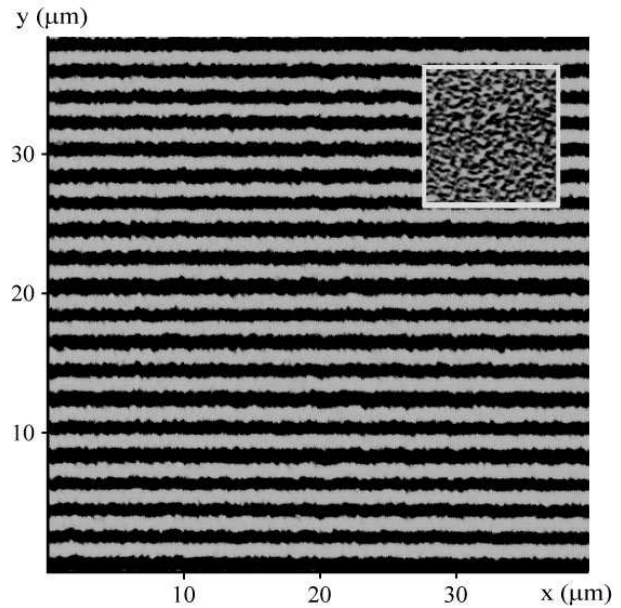}
} \caption{Magnetic force microscope image showing part of a pattern
of parallel lines written on the film. The dark regions are
unchanged from the original magnetisation, whereas the bright
regions show where the magnetisation is reversed. \textit{Inset}:
a nearly demagnetised piece of the Co/Pt film, showing the size of
the domains.} \label{fig:lines}
\end{figure}

Our first pattern is a set of parallel lines, as shown in
Fig.\,\ref{fig:lines}. With the reversed magnetic field set at
50\,mT and the laser power at 75\,mW, the sample is translated at
$250\,\mu$m/s along the $x$ axis. This produces a line of reverse
magnetisation $1\,\mu$m wide, which can be as long as the size of
the film. At the end of each line, the shutter is closed and the
sample is moved back to the start before making a step of
2\,$\mu$m along $y$ to the start of the next line.

The particular array shown in Fig.\,\ref{fig:lines} covered a
$200\,\mu$m square area. The figure shows a typical small section
of the array imaged by a magnetic force microscope.  The dark
regions indicate areas magnetised in the original direction whilst
the bright regions have reverse magnetisation. Such images give
quantitative information about the positions and widths of the
lines. They also allow us to see when a region is magnetically
saturated. If the film is demagnetised, the domains are mixed,
with some magnetised up and others down as shown inset in the top
right hand corner of Fig.\,\ref{fig:lines}. By contrast, the
uniformity of the lines in Fig.\ref{fig:lines} shows that their
magnetization is saturated. Along the boundaries between lines,
one can see small irregularities: the lines waver by approximately
250\,nm peak-to-peak. The period of this noise is typically
300\,nm but the spectral distribution is broad. We have not been
able to achieve straighter boundaries and we presume that lines
written in this way have a fundamental noise level imposed by the
domain size.

This method of writing only works for a relatively narrow range of
parameters. For example, if the laser power is lowered to 70\,mW,
there is hardly any magnetic response to the writing procedure.
However, if the scanning speed is reduced at the same time, the
film responds once again.  This seems reasonable as it must surely
be the temperature reached in the magnetic film that is the
essential parameter. If instead we increase the laser power to
80\,mW, we start to see physical damage to the surface of the
film. Atomic force microscope measurements show a 2\,nm high ridge
along the line drawn by the laser at this power. This can be
avoided by making a compensating change to higher scan rate. There
are also constraints on the strength of the reverse magnetic field
bias. Below 50\,mT, we start to see domain structure in the body
of the lines indicating that the remagnetisation is not saturated.
Above 100\,mT the domain irregularities along the boundary lines
begin to grow. For this reason, we operate at the lowest reverse
field that still saturates the body of the lines.

\subsection{Painting an area}

\begin{figure}
\resizebox{0.45\textwidth}{!}{
  \includegraphics{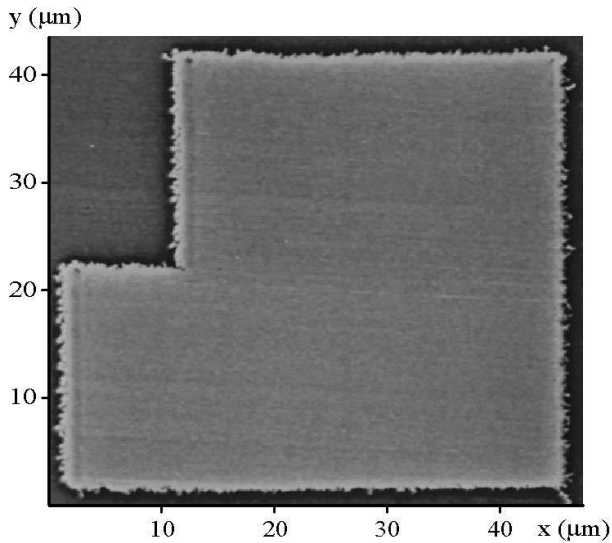}
} \caption{Magnetic force microscope scan of a Z-trap pattern. The
grey-scale in this image is identical to that in
Fig.\,\ref{fig:lines}} \label{fig:Z-trap}
\end{figure}

Our second pattern involves``painting'' an area of reversed
magnetisation in the shape of a 40\,$\mu$m square  with a piece
missing from the top left, as shown in Fig.\,\ref{fig:Z-trap}. The
motivation for making this particular shape is discussed in the
next section. Once again the pattern is made by drawing a series
of lines at $250\,\mu$m/s along the $x$ direction, but in this
case the step along $y$ is $0.5\,\mu$m, which makes successive
lines overlap. In order to achieve full overlap of the lines we we
used a higher laser power of 79\,mW and increased the reverse bias
field to 0.1\,T.

The edges parallel to the $x$ axis have a similar level of noise
to that seen in Fig.\,\ref{fig:lines}. However, the transverse
deviations can be as large as $1\,\mu$m peak-to-peak on the edges
parallel to $y$. Here the translation stage has decelerated as it
approaches the end of the line, resulting in an excess deposition
of energy. In combination with the larger reverse bias, this
promotes domain growth, leading to the protuberances seen in the
picture. We anticipate that this can be remedied by more precise
control over the shutter timing, which would allow us to start and
end the lines while the stage is still moving at constant
velocity. Further, improved actuation along the $y$ direction
would allow us to use a shorter step size, so that the pattern
could be written with a lower bias field and less laser power. We
believe that the noise along the edges of the pattern can be
reduced at least to the same level as seen for the array of lines
with these modifications of the apparatus.
\section{Application to atom microtraps}
\label{sec:analysis}

In thinking about the kinds of fields that could be made above the
surface, it is useful to replace the magnetisation by an
equivalent current density $\nabla\times M$\,\cite{Jackson1998}.
Along a boundary where the magnetisation reverses, going from $M$
to $-M$, this current density results in an equivalent current of
$2Mt$, where $t$ is the thickness of the magnetised material.
Taking our value of 2.3\,T/$\mu_0$ for the cobalt magnetisation
and 6\,nm for the thickness of the cobalt, we obtain an equivalent
current of 22\,mA along boundaries where the magnetisation
reverses. Thus, the structures we can make in this magnetic film
are equivalent to any network of 22\,mA current loops in a plane.

\begin{figure}
\includegraphics*[width=0.95\columnwidth]{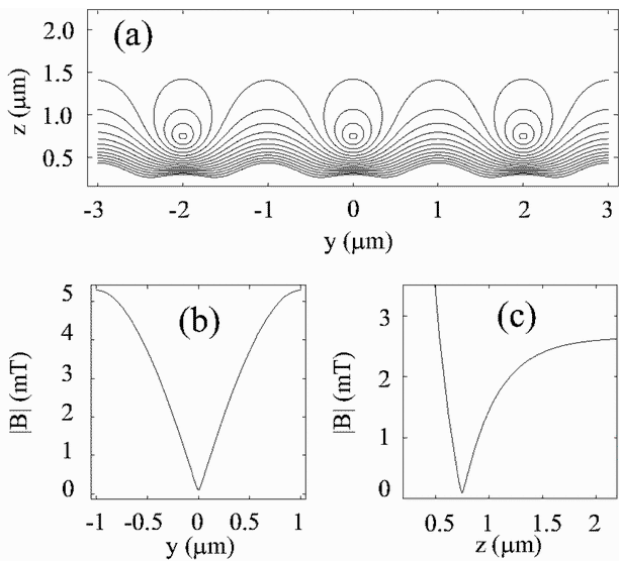}
\caption{Calculated magnetic field magnitude above an array of
lines (see Fig.\,\ref{fig:lines}) with a 2.6\,mT bias along $y$ to
form the traps and 0.1\,mT along $x$ to preserve the atomic spin
orientation. (a) Contours of constant field strength. (b) Field
strength versus $y$ through the centre of a trap at
$z=0.75\,\mu$m. (c) Field strength versus $z$ through the centre
of a trap at $y=0$.} \label{fig:linecurves}
\end{figure}

Viewed in this way, the pattern shown in Fig.\,\ref{fig:lines} is
equivalent to lines of current parallel to the $x$ axis, spaced by
$1\,\mu$m and alternating in direction. The magnetic field they
produce is easily calculated using the Biot-Savart law. A similar
pattern with 3\,$\mu$m period was recently fabricated by etching a
hard disk drive\,\cite{Lev2003}.  Larger structures of this kind
have previously been fabricated on videotape and used as atom
mirrors\,\cite{OurReview} and gratings\,\cite{Rosenbusch2000}.
Here, however, we are primarily interested in the application to
trapping because of the unusually high trap frequencies that can
be reached in small-scale structures. To be specific, let us
consider adding a uniform bias field of 2.6\,mT along the $y$
direction to interfere with the field created by the film.
Fig.\,\ref{fig:linecurves}(a) shows the resulting contours of
constant field strength. The circles in the contour plot enclose
lines of zero magnetic field that are formed by destructive
interference at a height $z$ of 0.75\,$\mu$m in a 2\,$\mu$m-period
array. Weak-field-seeking atoms may be trapped on these lines. An
atom held in one of these traps will lose its spin orientation if
the field is allowed to go strictly to zero, so we include here a
small uniform field $B_0=0.1\,$mT along the $x$ direction to
maintain the quantization.

Figs.\,\ref{fig:linecurves}(b) and (c) show the field strength
versus $y$ and $z$. In both cases this varies linearly in the
vicinity of the trap with a gradient of $B'=8.5\times 10^3$\,T/m,
except in the central $10-20$\,nm, where the bias $B_0$ is
significant. There the field strength grows quadratically, leading
to a transverse oscillation frequency for the trapped atom given
approximately by
\begin{equation} f = \frac{1}{2 \pi}\sqrt{{\mu \over m} {B'^2
\over B_0}}, \label{eq:trapfrequency}
\end{equation}
where $\mu$ and $m$ are the magnetic moment and mass of the atom.
For $^{87}$Rb atoms in the $F=2$, $m_F=2$ state, the trap in this
example has a very high oscillation frequency of $f=1.1$\,MHz.

We turn now to the $x$ direction, along the axis of the trap.
There is no trapping force in this direction, but axial
confinement can be achieved if necessary by additional
current-carrying wires underneath the chip. For example, two wires
positioned at $x=\pm 1000\,\mu$m, $z=-750\,\mu$m, each carrying
10\,A along $y$, will confine rubidium atoms in the axial
direction with an oscillation frequency of 82\,Hz. This is the
type of arrangement normally used in our group\,\cite{Jones2004}.

A more accurate description of the field above the film must take
into account the domain structure seen in Fig.\,\ref{fig:lines}.
Thus, the equivalent currents do not really flow along the $x$
direction but oscillate from side to side in the $xy$ plane.
 The typical amplitude is $a=125$\,nm and the typical wavelength is
$\lambda=300$\,nm. At height $z>>a$ above such a wire, the main
field component is
\begin{equation} B_y\simeq \mu_0 I/2\pi z\,, \label{eq:By}
\end{equation}
whilst the transverse oscillations produce an oscillating field
along the wire given by \cite{Jones2004}
\begin{equation} B_x\simeq \frac{\mu_0 I}{2\pi} k^2 a K_1(k z)\cos(k x)\,,
\label{eq:Bx}
\end{equation}
where $k=2\pi/\lambda$ and $K_1$ is the modified Bessel function.
It is this latter field component that makes the traps uneven and
causes atom clouds to break up above current-carrying wires. In
the present case of Rb atoms 750\,nm above the film, this field
has an amplitude of $\sim 10\,$nT, corresponding to an undulating
potential of $\sim\,10$nK. For many applications this amount of
roughness is perfectly acceptable, but for the most sensitive
experiments on quantum gases and quantum information processing it
is not.  We believe that further improvement would be worthwhile
and could be achieved by having smaller domain sizes. Alternative
writing procedures are also under investigation.

\begin{figure}
\includegraphics*[width=0.95\columnwidth]{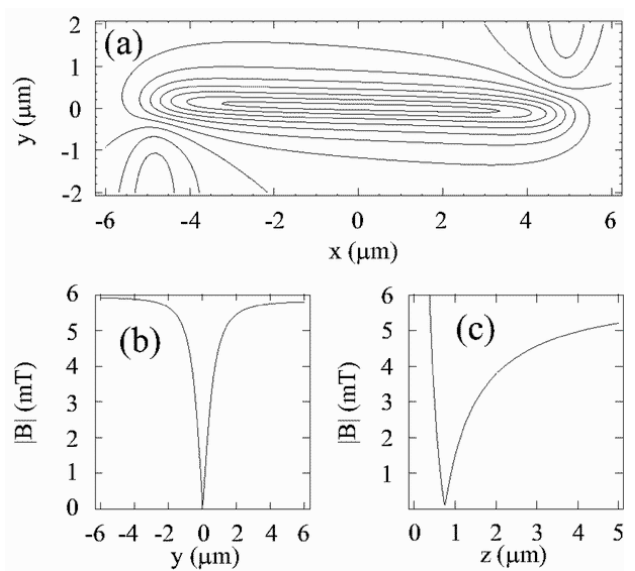}
\caption{Calculated magnetic field strength near the Z-trap of
Fig.~\ref{fig:Z-trap} with 5.8\,mT bias along y to form the trap.
(a) Contours of constant field strength. (b) Field strength versus
$y$ through centre of trap at $z=0.75\,\mu$m. (c) Field strength
through trap centre versus $z$.} \label{fig:Z-trapcurves}
\end{figure}

 For the magnetic pattern shown in
Fig.~\ref{fig:Z-trap}, the equivalent current is a single loop
encircling the region of reversed magnetisation. The Z-shaped edge
on the left is the region of interest for atom trapping purposes.
The short central section along $x$ is the ``wire'' above which
atoms are to be trapped, whilst the longer legs serve as the end
wires, providing both axial confinement\,\cite{ReichelPRL1999} and
a suitable bias of 0.12\,mT along $x$ for avoiding spin flip
transitions.  For a 22\,mA current flowing in this way, we
calculate that a bias field of 5.8\,mT along $y$ makes a trap
lying $0.75\,\mu$m above the film. Fig.~\ref{fig:Z-trapcurves}(a)
shows the contours of constant magnetic field strength in the $xy$
plane $0.75\,\mu$m above the film, with the origin defined to be
at the centre of the trap. Curves (b) and (c) show cuts through
the trap in the $y$ and $z$ directions. The transverse oscillation
frequency along $y$ or $z$ is 0.9\,MHz. In the axial direction
along $x$ it is 19\,kHz and in this case no auxiliary wires are
needed. If a more anisotropic trap is required, one simply has to
make the central section of the Z-shaped boundary be longer.

\section{Summary}
\label{sec:discussion}

We have shown that areas and lines of saturated magnetisation can
be written thermomagnetically on Co/Pt thin films. We have
illustrated the method with two patterns that are relevant for
producing high frequency neutral atom traps: the long wire and the
Z-trap. With these drawing and painting tools any pattern of
interest can be made at a resolution down to approximately
$1\,\mu$m. This small scale gives access to extremely tight traps
with MHz frequencies for rubidium atoms, making the method
promising for studies of 1D quantum gases and for small atom trap
arrays that could be suitable for quantum information processing.

In atom chips based on current-carrying wires, the magnetic field
close to a wire suffers from rf interference and thermal noise,
which cause atom loss through vibrational excitations and spin
flips. These effects should be avoided with the use of the Co/Pt
thin films. A further problem with current-carrying wires is that
the current wanders from side to side, leading to uneven traps. A
similar effect is found in the Co/Pt thin films due to the domain
structure. However the amplitude and period of these excursions
are small compared with a micron, making the effect negligible in
all but the tightest traps, i.e. those that are closest to the
surface.

\begin{acknowledgments}

We are indebted to Jon Dyne for expert technical assistance. We
also thank David Lau and Boris Vodungbo for their contributions to
the early stages of the project. This work was supported by the UK
Engineering and Physical Sciences Research Council and by the
FASTNET network of the European Union.

\end{acknowledgments}

\end{document}